\documentclass[namedreferences]{SolarPhysics}
\usepackage[optionalrh]{spr-sola-addons} 
\usepackage[pdftex]{graphicx}
\usepackage{graphicx}        
\usepackage{color}           
\usepackage{url}             

\begin{document}

\begin{article}

\begin{opening}

\title{Sources of SEP Acceleration during a Flare--CME Event}

\author{N.J.~\surname{Lehtinen}\sep
        S.~\surname{Pohjolainen}\sep
        K.~\surname{Huttunen-Heikinmaa}\sep
        R.~\surname{Vainio}\sep
        E.~\surname{Valtonen}\sep
        A.E.~\surname{Hillaris}
       }
\runningauthor{N.J. Lehtinen {\it et al.}}
\runningtitle{Sources of SEP Acceleration}

   \institute{N.J. Lehtinen, S. Pohjolainen;  
                     Tuorla Observatory/Department of Physics, 
                     University of Turku, 21500 Piikki\"o, Finland
                     email: \url{silpoh@utu.fi} \\
              K. Huttunen-Heikinmaa, E. Valtonen;
                     Department of Physics, University of Turku, 
                     20014 Turku, Finland\\
              R. Vainio; 
                     Department of Physical Sciences, PO Box 64, 
                     00014 University of Helsinki, Finland\\
              A.E. Hillaris;  
                     Department of Physics, University of Athens, 
                     15784 Panepistimiopolis Zografos, Athens, Greece
                                   }

\begin{abstract}
A high-speed halo-type coronal mass ejection (CME), associated with 
a GOES M4.6 soft X-ray flare in NOAA AR 0180 at S12W29 and an EIT 
wave and dimming, occurred on 9 November 2002. A complex radio event was 
observed during the same period. It included narrow-band fluctuations 
and frequency-drifting features in the metric wavelength range, 
type III burst groups at metric--hectometric wavelengths, 
and an interplanetary type II radio burst, which was visible 
in the dynamic radio spectrum below 14 MHz. 
To study the association of the recorded solar energetic particle (SEP)
populations with the propagating CME and flaring, we perform 
a multi-wavelength analysis using radio spectral and imaging 
observations combined with white-light, EUV, hard X-ray, and 
magnetogram data.  
Velocity dispersion analysis of the particle distributions (SOHO and 
{\it Wind} {\it in situ} observations) provides estimates for the release 
times of electrons and protons. Our analysis indicates that proton 
acceleration was delayed compared to the electrons.    
The dynamics of the interplanetary type II burst identify the burst 
source as a bow shock created by the fast CME.
The type III burst groups, with start times close to the estimated 
electron release times, trace electron beams travelling along 
open field lines into the interplanetary space.  
The type III bursts seem to encounter a steep density gradient 
as they overtake the type II shock front, resulting in an abrupt 
change in the frequency drift rate of the type III burst emission.
Our study presents evidence in support of a scenario in which 
electrons are accelerated low in the corona behind the CME shock 
front, while protons are accelerated later, 
possibly at the CME bow shock high in the corona. 
\end{abstract}
\keywords{Coronal Mass Ejections, Initiation and Propagation; 
Energetic Particles, Electrons, Protons; Radio Bursts, Meter-Wavelengths 
and Longer}
\end{opening}

\section{Introduction}

Coronal mass ejections (CMEs) are large-scale structures carrying plasma and 
magnetic field from the Sun. CMEs are associated with flares, filament 
eruptions, shocks, radio bursts, solar energetic particle (SEP) events, and 
have been found to be the primary cause of major geomagnetic disturbances and 
changes in the solar wind flow (see, {\it e.g.}, Schwenn {\it et al.},  
\citeyear{schwenn05}; Vilmer {\it et al.}, \citeyear{vilmer03}; Mann 
{\it et al.}, \citeyear{mann03}; Gopalswamy {\it et al.}, 
\citeyear{gopalswamy02}; Leblanc {\it et al.}, \citeyear{leblanc01}; Gosling, 
\citeyear{gosling97}).

Radio type III bursts are fast frequency-drifting features that can be used 
as tracers of energetic electrons as they are caused by subrelativistic 
($\approx$2.5 to $\approx$80 keV) electron streams travelling outward from 
the Sun at speeds of about 0.1\,c to 0.5\,c 
(see, {\it e.g.}, Wild, \citeyear{wild50}; 
Dulk, \citeyear{dulk85}). Near-relativistic electrons (30\,--\,315 keV) can be 
observed {\it in situ} during SEP events, in association with type III 
bursts that continue into the IP space (Simnett \citeyear{simnett05}). 
Radio type II bursts, 
on the other hand, are thought to be caused by MHD shock waves 
propagating through the corona and interplanetary medium ({\it e.g.}, 
Nelson and Melrose, \citeyear{nelson85}). Some studies suggest
that coronal metric type II bursts and interplanetary (IP) 
decameter--hectometer (DH) type II bursts do not have the same 
driver ({\it e.g.}, Cane and Erickson, \citeyear{cane05}), 
and that the latter are exclusively created by the CME bow 
shocks ({\it e.g.}, Cliver, Webb, and Howard, \citeyear{cliver99}).

SEPs are believed to be accelerated in the lift-off phase of CMEs, which could 
include either acceleration in flaring processes ({\it e.g.}, de Jager,  
\citeyear{jager86}), acceleration in coronal/interplanetary shocks 
({\it e.g.}, Reames, \citeyear{reames99} and references therein), or  
in both ({\it e.g.}, Klein and Trottet, \citeyear{klein01}; Klein {\it et al.}, 
\citeyear{klein99}; Kocharov and Torsti, \citeyear{kocharov02}). Recently, the 
role of CME interaction in SEP production has also been discussed ({\it e.g.}, 
Gopalswamy {\it et al.}, \citeyear{gopalswamy02}; Kahler, \citeyear{kahler03}; 
Wang {\it et al.}, \citeyear{wang05}, and references therein). One of the 
difficulties in determining the relevant processes lies in the 
uncertainties of timing. SEPs are typically released no 
earlier than the maxima of the flare impulsive phases 
(Kahler, \citeyear{kahler94}), yet at the time of SEP release 
the height of the associated CME can vary from 
one to over ten solar radii (Kahler, \citeyear{kahler94}; 
Krucker and Lin, \citeyear{krucker00}; Huttunen-Heikinmaa, Valtonen, 
and Laitinen, \citeyear{huttunen05}). It has also been observed that 
different particle species can be released non-simultaneously 
(Huttunen-Heikinmaa, Valtonen, and Laitinen, \citeyear{huttunen05}; 
Krucker and Lin, \citeyear{krucker00}; Mewaldt {\it et al.}, 
\citeyear{mewaldt03}).

The {\it in-situ} observations of SEPs are strongly influenced by the magnetic 
connection between the spacecraft and the flare site ({\it e.g.}, Cane, 
Reames, and von Rosenvinge, \citeyear{cane88}). For example, the 
nominally well-connected area for the spacecraft in the L1-point is 
around 60$^{\circ}$ West on the solar hemisphere. 
\inlinecite{kallenrode93} found that at larger angular distances 
between the observer's magnetic footpoint and the flare location, the onsets 
of energetic particles in relation to the microwave maximum are more 
delayed, and the proton onsets are relatively more delayed than the electron 
onsets.  Assuming that SEPs originated from the flaring processes, 
\inlinecite{kallenrode93} concluded that the propagation of $\approx$20 MeV 
protons from the flare site to the connecting magnetic field line is 
systematically slower than of $\approx$0.5 MeV electrons, and the 
electron-to-proton ratio tends to increase with angular distance from 
the flare site. Of course, the modern view of particle acceleration in 
the solar corona does not rely on flares as the only source of energetic 
particles during SEP events. Thus, instead of reflecting the azimuthal 
particle propagation conditions in the corona, the findings of Kallenrode 
can be re-interpreted as reflecting the azimuthal and/or temporal 
evolution of the source, usually a coronal shock wave driven/launched 
by a CME;
a delay between the plasma eruption and the observed onset of the SEP event 
reflects the establishment of the magnetic connection between the source 
and the observer and/or the efficiency of the acceleration and escape 
of particles in different parts (bow/flanks) of the global coronal 
shock wave. 

Consistent with the current views and the conclusions of 
\inlinecite{kallenrode93} are the results of \inlinecite{krucker99}, 
who found that in some events the injection time of the electrons is 
consistent with the timing of the type III radio emission, while in 
other cases the electrons were apparently released up to half an hour 
later than the start of the type III burst. The non-delayed electron 
events were concentrated in the well-connected area, while the delayed 
events were more spread out. Similar delays were also observed between 
protons and helium nuclei by \inlinecite{huttunen05}. The 
near-simultaneous proton and helium events originated from the 
well-connected area, and the events with delayed helium nuclei were 
more spread out.

In this paper we study the possible SEP accelerators on 9 November 2002. 
In Section 2 we analyse the multiwavelength observations of this event: 
in 2.1 we describe how SEP onset times were estimated, in 2.2 we analyse 
the flaring activity and CMEs near the SEP onset times, and in 2.3 we 
identify accelerated particles via radio emission. In Section 3 we 
calculate heights for the radio features and compare them with other
structures. Results are presented in Section 4 and they are further 
discussed in Section 5.

\section{Observations and Data Analysis}

The energetic particle data for protons were recorded by SOHO/ERNE 
\cite{torsti95}, and for electrons by {\it Wind}/3DP \cite{lin95} and 
SOHO/EPHIN \cite{muller95}. 
White-light coronograph images were provided by SOHO/LASCO 
\cite{brueckner95}, EUV images by SOHO/EIT \cite{delaboud95},
hard X-ray data by RHESSI \cite{lin02}, and solar magnetograms by 
SOHO/MDI \cite{scherrer95}. 
The radio imaging at meter wavelengths was done by the Nan\c{c}ay 
Radioheliograph in France (NRH; Kerdraon and Delouis, \citeyear{kerdraon97}). 
Radio spectral data at decimetric--metric wavelengths were
provided by Artemis-IV in Greece \cite{kontogeorgos06}, Phoenix-2 
in Switzerland \cite{messmer99}, and Nan\c{c}ay Decameter Array in France 
(DAM; Lecacheux, \citeyear{lecacheux00}). At decameter--hectometer 
wavelengths the radio spectral data were obtained from 
the WAVES experiment \cite{bougeret95} onboard the 
{\it Wind} satellite.

\subsection{Energetic Particles}

An observable signature of SEP events is a velocity dispersion. Assuming that 
particles with different energies are released simultaneously at or close to 
the Sun, the onset of the event at a distance from the Sun should be observed 
earlier at higher energies than at lower ones. Velocity dispersion analysis 
makes it possible to infer the release time of particles at the Sun and the 
apparent path length travelled. 
The velocity dispersion analysis has been utilized by several
authors ({\it e.g.}, 
Debrunner, Flueckiger, and Lockwood, \citeyear{debrunner90}; 
Torsti {\it et al.}, \citeyear{torsti98}; 
Krucker {\it et al.}, \citeyear{krucker99}; 
Mewaldt {\it et al.}, \citeyear{mewaldt03}; 
Tylka {\it et al.}, \citeyear{tylka03}; 
Dalla {\it et al.}, \citeyear{dalla03}; 
Hilchenbach {\it et al.}, \citeyear{hilch03}). Recently, the validity of the 
velocity dispersion analysis has been examined by simulation studies 
(Lintunen and Vainio, \citeyear{lintunen04}; 
Saiz {\it et al.}, \citeyear{saiz05}).

In this study, we determine the {\it in-situ} onset times of SEP events 
by adopting the method used by \inlinecite{huttunen05}. The 
release times at the Sun are then first derived using the velocity dispersion 
analysis. However, since the analysis has several complex systematic errors, 
we finally estimate the temporal windows for the particle release at the Sun 
using a relaxed condition for the path length. 
For more detailed discussions on deriving the release times at the Sun, we
refer to \inlinecite{lintunen04} and \inlinecite{huttunen05}, and 
references therein.

The velocity dispersion equation is 
\begin{equation}
t_{onset}(E)=t_0+\frac{s}{v(E)},
\end{equation}
where $t_{\rm onset}(E)$ is the observed onset time at energy E at the
spacecraft, $t_0$ is the release time from the acceleration site, 
$s$ is the apparent path length travelled by the particles, 
and $v(E)$ is the velocity of the particle of energy $E$. The parameters 
are determined by fitting a straight line to the points 
$\{1/v(E_{j}),~ t_{{\rm onset}}(E_{j})\}$. 
To make the release time derived from Equation (1) comparable to 
the electromagnetic observations near Earth, eight minutes have 
to be added ($t_{0\ast}$=$t_0$ + 8 min).

\begin{figure}[!ht]
\centering
  \includegraphics[width=1.0\textwidth]{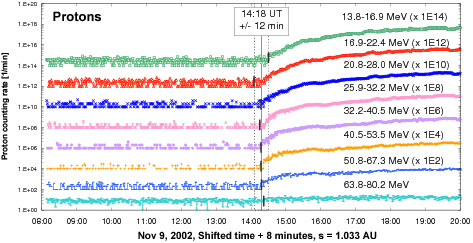}
\caption{The proton count rates observed by SOHO/ERNE. 
The profiles are shifted back in time assuming the path length 
of 1.03 AU, acquired from the velocity dispersion fit. The thick short
vertical lines represent the onset times on different channels, and the
proton release time close to the Sun is marked by a long vertical solid
line (dashed vertical lines show the error limits). For visual reasons,
the profiles have been scaled suitably, and eight minutes have 
been added to make them comparable to the electromagnetic observations.
}
\label{fig:1}       
\end{figure}

The velocity dispersion analysis of the SOHO/ERNE proton data in the 
14\,--\,80 MeV energy range yielded a path length of 1.03\,$\pm$\,0.33 AU. 
This energy range is covered entirely by the High Energy Detector (HED) 
of the ERNE instrument.  With this path length the proton acceleration 
was estimated to have taken place at 14:18 UT $\pm$ 12 minutes.
The proton count rates, at several energy channels shifted back 
in time to compensate for the velocity dispersion, are presented in 
Figure \ref{fig:1}. 

\begin{figure}[!h]
\centering
  \includegraphics[width=1.0\textwidth]{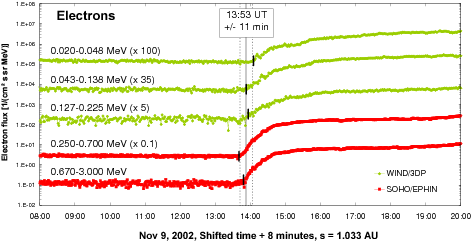}
\caption{The shifted electron intensity profiles based on observations 
by SOHO/EPHIN and {\it Wind}/3DP. The applied path length is the 
same as for protons (1.03 AU). See Figure \ref{fig:1} for details.}
\label{fig:2}       
\end{figure}

\begin{figure}[!h]
\centering
\includegraphics[width=0.7\textwidth]{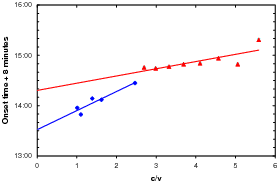}\\  
\vspace{5mm}
\includegraphics[width=0.7\textwidth]{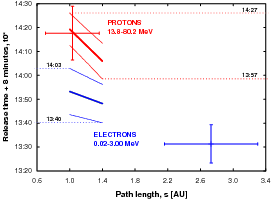}
\caption{
Fits of the velocity dispersion equation to proton and
electron data (top panel). Proton data (triangles) are from the ERNE
High Energy Detector HED and electron data (diamonds) from the 
{\it Wind}/3DP and SOHO/EPHIN. Bottom panel shows the estimated release time 
windows of protons and electrons. Thick inclined lines
represent the average proton (red) and electron (blue) release times, 
when the onset times of several energy channels are shifted back in 
time with path lengths of 1.0\,--\,1.4 AU. The time windows (horizontal 
dashed lines) are based on one standard deviation error limits of 
the average proton and electron release times (thin inclined lines). 
The solid crosses represent the results with statistical error limits 
when the velocity dispersion analysis is carried out separately for 
proton and electron data.
} 
\label{fig:3}       
\end{figure}

Using {\it Wind}/3DP (0.020\,--\,0.225 MeV) and SOHO/EPHIN 
(0.25\,--\,3.00 MeV) {\it in situ} electron observations and assuming 
the same path length as for the protons, the electron acceleration is 
estimated to have started at 13:53 UT $\pm$ 11 minutes (Figure \ref{fig:2}). 
However, the path length of 1.03 AU seems to fail in compensating the 
velocity dispersion of the electron onsets. The velocity dispersion 
analysis performed only on the electron data yields a very high path 
length of 2.73\,$\pm$\,0.57 AU, which we deem somewhat unreliable 
(Figure \ref{fig:3}). With this high path length the electron release 
times would have been earlier, at 13:31 UT $\pm$ 8 minutes, but most likely 
the high value of the path length is caused by an inaccurate 
determination of the onset times in the low-energy electron channels 
due to the long rise time and high background level of the fluxes.
More conservative estimates of the release time windows are evaluated 
using path lengths in the 1.0\,--\,1.4 AU range. 

To evaluate the uncertainties in the release times, the standard
deviation error limits (1$\sigma$) for the average proton and electron
release times were calculated and used to determine the release time
windows, as shown in Figure \ref{fig:3}. It is clear that protons are 
delayed with respect to electrons. The delay is much larger if we 
accept the very high path lengths for the electrons obtained from the 
velocity dispersion fit to the electron data only.
Based on Figure \ref{fig:3}, we estimate that the protons in this 
event were released between 13:57 and 14:27 UT, and the electrons earlier, 
between 13:40 and 14:03 UT, assuming a low path length for both 
populations. The electrons could have been released even earlier 
than that, around 13:23\,--\,13:39 UT, if a much higher path length 
is accepted for the electron population.

\subsection{CMEs and Flares on 9 November 2002}

\begin{figure}[!h]
\centering
  \includegraphics[width=1.0\textwidth]{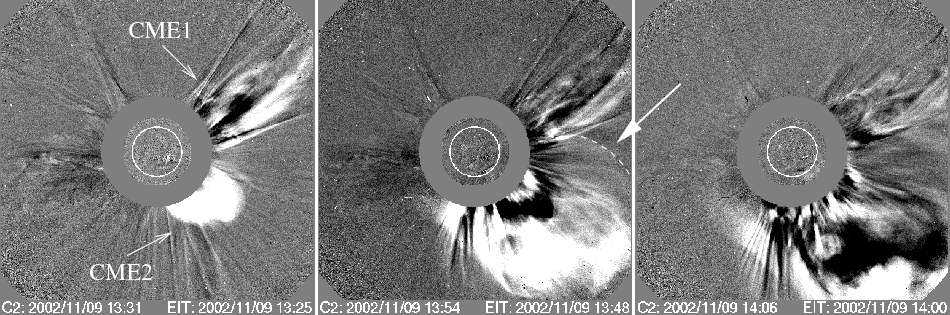}
\caption{Evolution of the two CMEs observed on 9 November 2002. 
The first, directed to the Northwest (CME1), had a plane-of-the-sky 
speed of about 530 km s$^{-1}$. The later halo-type CME (CME2), 
was directed to the Southwest and had a plane-of-the-sky speed of 
about 1840 km s$^{-1}$. The white dashed line (indicated with an arrow) 
outlines a brightening on the side of the halo CME at 13:54 UT. }
\label{fig:4}       
\end{figure}

On 9 November 2002, during the estimated SEP release period, two CMEs 
were observed. The first (CME1 in Figure \ref{fig:4}) was directed 
to the Northwest. It was first detected at 11:06 UT in LASCO C2, 
at a height of 2.55 R$_\odot$, travelling at a plane-of-the-sky speed of 
530 km s$^{-1}$ (determined from a linear fit to all height-time 
data points from the LASCO CME Catalog). 
The second CME (CME2 in Figure \ref{fig:4}), was halo-type and  
directed to the Southwest, and it was first detected at 13:31 UT at 
a height of 4.20 R$_\odot$. The plane-of-the-sky speed from a linear 
fit to all data points was 1840 km s$^{-1}$. A second-order fit to the 
same data points indicates that the halo CME was accelerating, reaching 
a speed of 2000 km s$^{-1}$ near 30 R$_\odot$. We estimate that the lateral 
expansion speed (see Schwenn {\it et al.}, \citeyear{schwenn05}) was
even higher than that, about 2300 km s$^{-1}$ before 14:00 UT, and about 
1800 km s$^{-1}$ after 14:00 UT. This speed is, however, difficult to 
measure, since the northwestern CME2 edge was irregular and varying 
in time. The three consecutive LASCO C2 difference images in 
Figure \ref{fig:4} show how the two CMEs evolved and how the halo 
CME expanded laterally. At 13:54 UT (middle panel in Figure \ref{fig:4}) 
an arc-like brightening is visible between the two CMEs. 
In the later LASCO C3 images the halo CME surrounds the disk in 
full 360 degrees.   

\begin{figure}[!h]
\centering
  \includegraphics[width=1.0\textwidth]{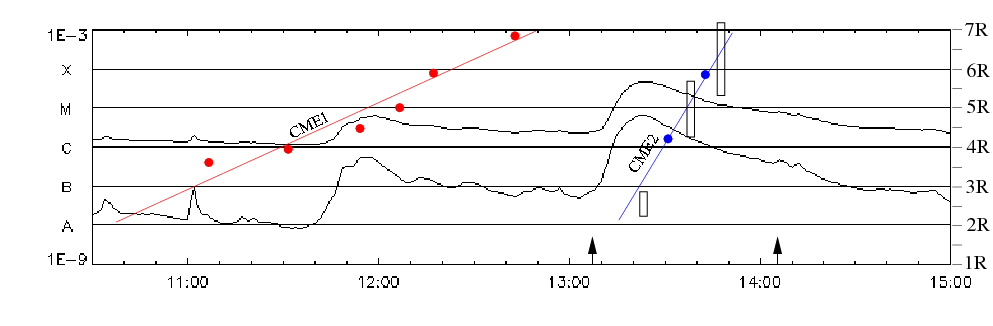}
\caption{A height--time plot of the plane-of-the-sky leading fronts of 
the two CMEs (data from the LASCO CME Catalog) together with the GOES 
soft X-ray flux curve. Black arrows mark the start times of the 
RHESSI hard X-ray bursts at 13:07 and 14:05 UT. (Note that RHESSI 
did not observe during the satellite night-time, between 13:28 and 
14:00 UT). Estimated heights for the DH type II burst source are 
also shown, boxes represent the calculated heights according to the
Saito and hybrid density models, see Section 3 for details. 
}
\label{fig:5}       
\end{figure}

\begin{figure}[!h]
\centering
  \includegraphics[width=0.75\textwidth]{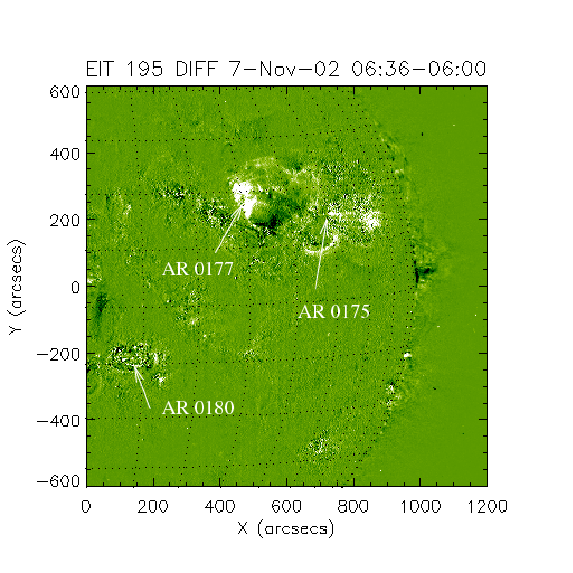}
  \includegraphics[width=0.55\textwidth]{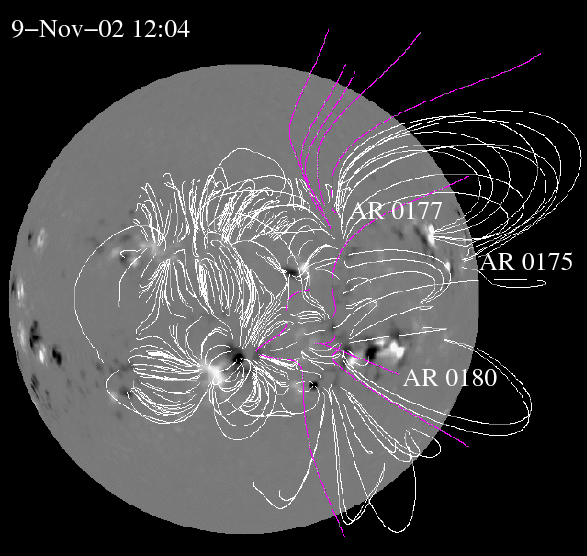}
\caption{Top: Active regions associated with the analysed 
flares and CMEs, observed by SOHO/EIT on 7 November 2002 (difference 
image at 06:36\,--\,06:00 UT). By 9 November these regions had rotated 
about 30$^{\circ}$ to the West.
Bottom: Potential field lines from magnetic field extrapolations (PFSS) 
using SOHO/MDI magnetogram on 9 November 2002 at 12:04 UT. Field 
lines plotted in magenta are open, with negative polarity.     
}
\label{fig:6}       
\end{figure}

The height--time plot in Figure \ref{fig:5} shows the estimated heights 
of the CME leading edges, combined with the GOES soft X-ray flux 
curve during 10:30\,--\,15:00 UT. The backward-extrapolated height--times 
for CME2 give an estimated start time for the halo CME near 13:05 UT.
The fast halo CME2 was clearly associated with activity in NOAA AR 0180 
at S11W36 (location on 9 November at 12:00 UT). The active regions most 
probably associated with the earlier CME1 were AR 0177 at N18W30 
(location on 7 November at 06:30 UT), and AR 0175 at N15W81 (location 
on 9 November at 15:30 UT ). These active regions are shown in an 
EIT map on 7 November, in Figure \ref{fig:6}. The difference in 
latitude between the source regions of the two CMEs was 25\,--\,30$^{\circ}$ 
and the difference in longitude 20\,--\,40$^{\circ}$, depending on the 
origin of CME1 (AR 0175 or AR 0177). 
Since CME2 developed into a full halo and showed lateral expansion, 
it is possible that the two CMEs made contact. 
However, as CMEs are observed as projections on the plane of the sky, 
determining the real geometry is prone to uncertainties. 

\begin{figure}[!h]
\centering
  \includegraphics[width=1.0\textwidth]{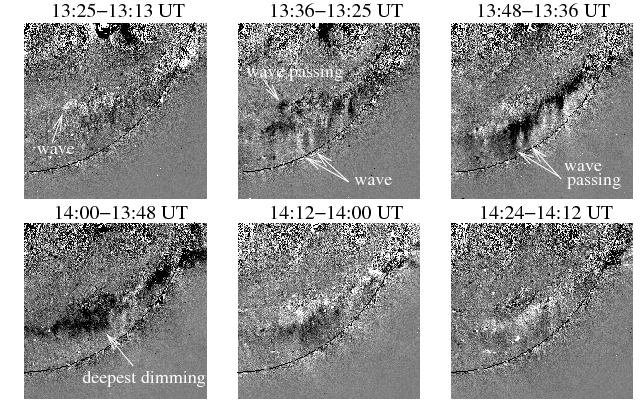}
\caption{SOHO/EIT 195 \AA \ difference images showing the EIT 
wave, wave passing, and dimmings. The shown area is $1000'' \times 800''$ 
in each frame.}
\label{fig:7}       
\end{figure}

Several GOES C-class flares were recorded during the first CME passage, 
before 13:00 UT on 9 November, but for many of them the location was not 
identified. At 13:08 UT a GOES M4.6 class flare started in AR 0180, 
at S12W29. The flux maximum occurred at 13:23 UT and the flux 
decayed to the pre-flare level at about 15:00 UT. 
The timing indicates clear association with the fast halo CME2.
RHESSI observed the flare in hard X-rays until 13:28 UT, when satellite 
night-time set in. The highest energy band in which the flare was observed 
was 100\,--\,300 keV. According to the RHESSI flare list the flare was 
located at $x\,=\,440''$, $y\,=\,-264''$ from the disk center. 
After the satellite night-time ended, RHESSI observed an X-ray 
brightening at 14:05 UT, slightly off from the earlier flare site, 
at $x\,=\,552''$, $y\,=\,-200''$.

The potential field ({\it i.e.}, the current-free field) extrapolations 
provide an approximation of the magnetic fields in a quiet Sun.
However, the lack of electric current and free energy in the model means 
that it does not describe active regions very well. The PFSS Solarsoft 
package \cite{derosa03} uses the line-of-sight magnetograms provided 
by SOHO/MDI, and the potential field  in Figure \ref{fig:6} 
describes the situation on 9 November. On the east side of AR 0180 
there are some open field lines (plotted in magenta, indicating 
negative polarity), and many more in the regions between AR 0180 
and AR 0177. It should be noted that the PFSS maps describe the
potential field prior to the event, and that ongoing events can strongly
perturbe the magnetic structures. 

In EUV, coronal loops were observed to open near the nortwest limb
at 09:48 UT, at the same latitude where AR 0175 and AR 0177 were located. 
An EUV brightening appeared in this region at 12:24 UT, and in the 
following EIT image bright material was ejected over the limb. These 
transients were observed well before the halo CME2 appeared. 

EUV ribbons were formed over the AR 0180 (S12W29) in the EIT image 
at 13:13 UT, and they were clearly associated with the GOES M4.6 flare. 
Post-flare loops appeared at 13:25 UT and an EIT wave was observed in
the same image. At first it showed as a bright ``rim'' South of AR 0180, 
and then as dark dimmed structures near the southwest limb after 
13:36 UT (Figure \ref{fig:7}). The dark structures persisted 
well after 14:12 UT. A small EUV brightening was observed in AR 0180 
at 14:12 UT, possibly associated with the RHESSI hard X-ray brightening 
at 14:05 UT.

\subsection{Radio Observations}

\begin{figure}[!hb]
\centering
  \includegraphics[width=0.9\textwidth]{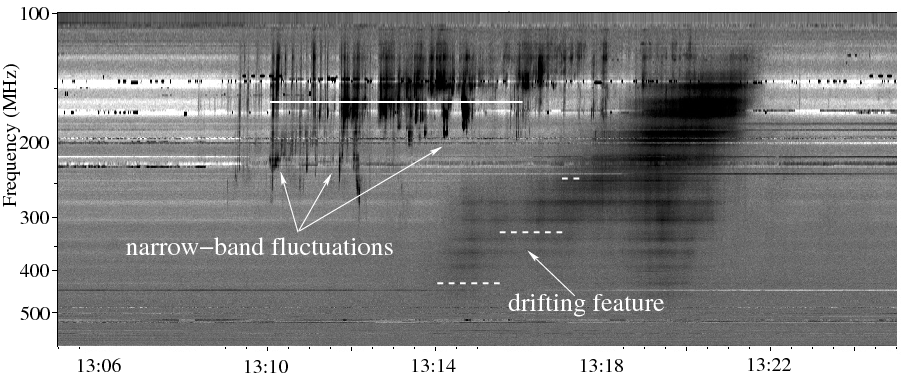}
\caption{Artemis-IV spectrum (100\,--\,600 MHz, 13:05\,--\,13:25 UT) shows 
narrow-band fluctuations and a frequency-drifting feature close to 
the start time of the GOES M4.6 flare. Figure \ref{fig:9} shows 
some of these source locations, imaged by NRH (times and frequencies 
are shown here with a white solid line and white dashed lines, 
respectively). After 13:17:30 UT the features become unclear due 
to a group of metric type III bursts superposed on the drifting 
emission.} 
\label{fig:8}       
\end{figure}

\begin{figure}[!ht]
\centering
  \includegraphics[width=0.9\textwidth]{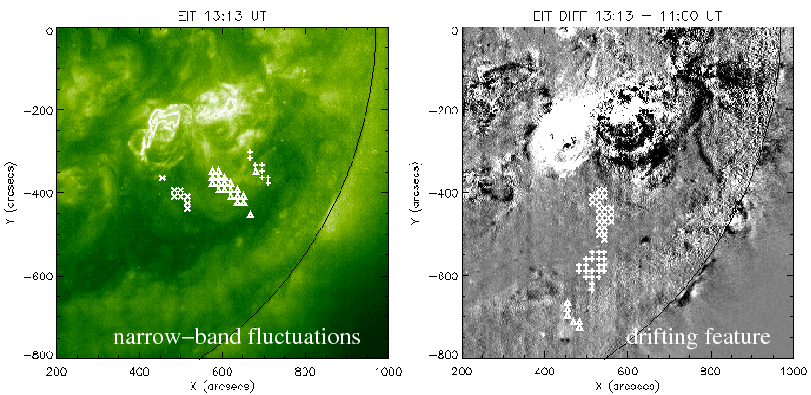}
\caption{Left: NRH radio source positions of the 
narrow-band fluctuations at 164 MHz, plotted over a SOHO/EIT 
image at 13:13 UT. Right: NRH radio source positions of the drifting burst 
feature, plotted over a SOHO/EIT difference image 13:13\,--\,11:00 UT. 
The emission drifts from 432 MHz (marked with $\times$), 
through 327 MHz (marked with +), to 236 MHz (marked with $\triangle$). 
The source positions indicate that the burst driver is moving 
southward from the AR, with a slight path curvature towards the East.
(The imaging periods were indicated in the dynamic spectrum 
in Figure \ref{fig:8}.)  
} 
\label{fig:9}       
\end{figure}

With radio observations we can trace accelerated particles and observe 
propagating shock fronts. We therefore studied radio emission during  
13:00\,--\,15:00 UT at a large wavelength range (4 GHz\,--\,100 kHz). 
The dynamic radio spectrum from Artemis-IV at 100\,--\,600 MHz shows 
narrow-band fluctuations (nbf) starting around 13:09 UT, one minute 
after the GOES M4.6 flare start. 
The fluctuations occured at metric wavelengths, in the 300\,--\,100 MHz 
frequency range. The fluctuating envelope drifted slowly towards the 
lower frequencies (Figure \ref{fig:8}). The emission source locations 
could be imaged by NRH at 164 MHz, and Figure \ref{fig:9} (left panel) 
shows how the emission centroids were located South of AR 0180, along 
and over closed loop structures. 

Another drifting feature (df) was observed to start around 13:14 UT  
just below 500 MHz, with a frequency drift around 0.8 MHz s$^{-1}$. 
The dashed lines in Figure \ref{fig:8} mark the frequencies that
could be imaged with NRH, and Figure \ref{fig:9} shows the source 
locations at the three different frequencies. The source trajectory 
indicates that the burst driver was moving southward from the AR, 
with a slight path curvature towards the East. We estimate from this
image that the projected speed of the emission source was around 
1200\,--\,1400 km s$^{-1}$.

\begin{figure}[!ht]
\centering
  \includegraphics[width=\textwidth]{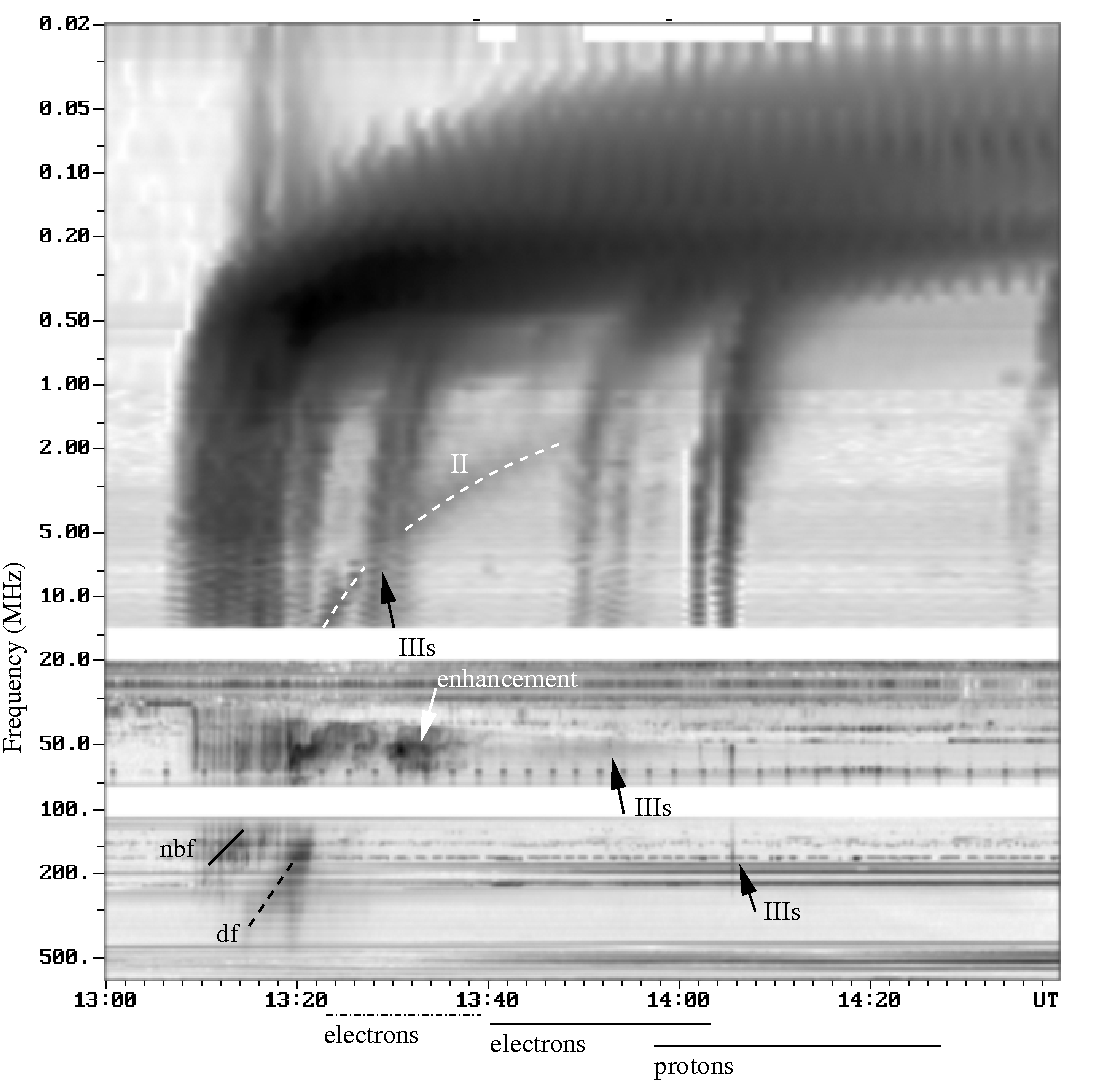}
\caption{
Composite of radio spectral data at 13:00\,--\,14:40 UT on 9 November 2002
(Artemis-IV and {\it Wind}/WAVES). The narrow-band fluctuations (nbf) 
and the drifting feature (df) are observed during 13:09\,--\,13:20 at
frequencies higher than 100 MHz (shown in detail in Figure \ref{fig:8}). 
Both of these features could have a continuation at frequencies lower
than 70 MHz. At 13:28\,--\,13:31 UT a type III burst group is observed 
at DH wavelengths, but the starting frequency is not clear. 
During 13:30\,--\,13:40 UT enhanced burst emission is visible 
at 35\,--\,65 MHz. Around 13:48 UT a group of type III bursts starts 
near 60 MHz, followed by another group around 14:05 UT around 250 MHz.
An IP type II burst becomes visible around 13:23 UT at 14 MHz. There is 
some indication that the burst could have started earlier at higher 
frequencies, but this emission is mixed with type III bursts. The type 
II emission lane can easily be followed down to 2 MHz near 14:00 UT, when 
the more intense type III bursts prevent seeing the fainter lane. 
Estimated electron and proton release times for both path lengths 
are also marked in the plot. Solid lines represent the release times 
using 1.0\,--\,1.4 AU path lengths and dash-dotted line path lengths 
$>$2.1 AU, see Section 2.1 for details.
}
\label{fig:10}       
\end{figure}

The{\it Wind}/WAVES observations at 14 MHz\,--\,20 kHz (Figure \ref{fig:10}) 
show an IP type II burst lane at $\approx$13:20 UT onwards. There is some 
indication that the burst starts at a frequency higher than the RAD2 
upper limit at 14 MHz, as we see emission lanes also around 30 MHz 
(ground-based observations). 
During 13:30\,--\,13:40 UT burst-like emission is observed also at 
35\,--\,65 MHz, and this enhanced emission looks to be separate from 
the type II burst lane. 
The type II emission drifts down to 2 MHz by 13:48 UT, when it 
disappears under the stronger IP type III burst groups. 
The type II burst lane appears occasionally also after 14:00 UT, and it 
is seen by WAVES RAD1 at 500 kHz near 15:20 UT. 

Several metric--DH type III bursts were observed in association with the 
main impulsive phase of the M4.6 flare. The first type III burst group 
that appeared in the estimated time-frame for particle acceleration  
was observed at 13:28 UT, but the starting frequency is overshadowed by other
burst structures. The next group of type III bursts started around 13:48 UT,
with starting frequency near $\approx$60 MHz (Figure \ref{fig:10}; this 
frequency was also verified from the DAM dynamic spectrum at 20\,--\,70 MHz, 
not shown here). Another type III burst group followed at 14:02 UT. Within 
this group burst lanes were observed to start at metric wavelengths 
near 250 MHz at $\approx$14:05 UT (frequency also verified from the 
Phoenix-2 spectrum).

\begin{figure}
\centering
  \includegraphics[width=1.0\textwidth]{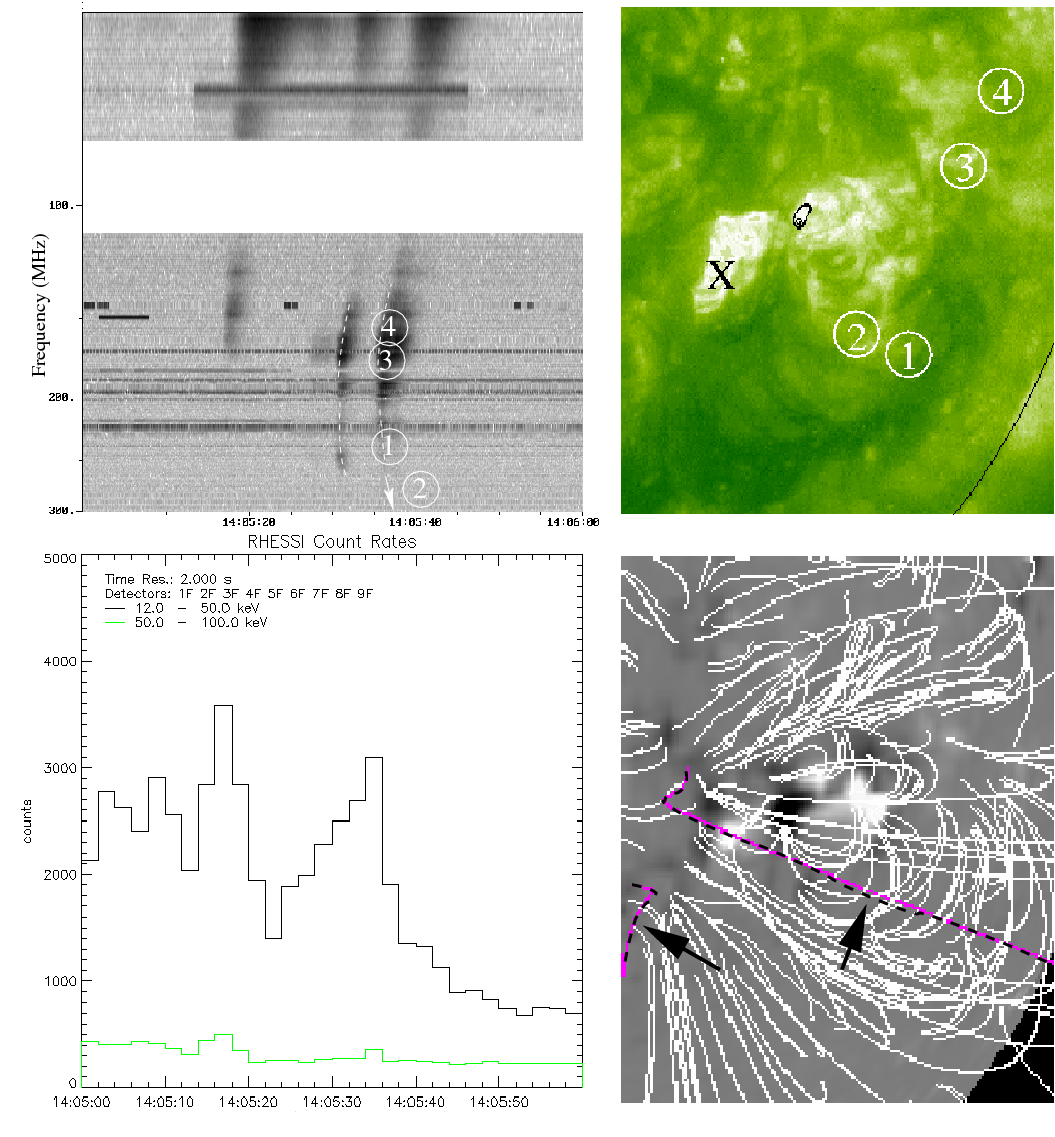}
\caption{Top left: Artemis-IV dynamic spectrum at 14:05:00\,--\,14:06:00 
UT at 50\,--\,300 MHz shows some of the type III bursts in detail. The 
times and frequencies of the imaged sources 1, 3, and 4, are indicated 
in the spectrum (burst source 2 at 327 MHz is outside the frequency 
range shown here). 
Top right: Emission locations of burst sources 1\,--\,4. X marks
the location of the GOES M4.6 flare with flare maximum at 13:23 UT. 
Contours show the location of the RHESSI hard X-ray burst at 
14:05:00\,--\,14:06:00 UT. 
Bottom left: RHESSI count rates from the same time period. 
Bottom right: Potential magnetic field (PFSS) in the same region as 
in the top right image, extrapolation is from the SOHO/MDI magnetogram 
at 12:04 UT on 9 November. Arrows point to the open field-line regions.
}
\label{fig:11}       
\end{figure}

On closer inspection, the type III bursts that started near 250 MHz 
look to be bidirectional. The burst emission drifts both upward and 
downward in frequency (Figure \ref{fig:11}). Bursts with a negative 
frequency-drift continue to DH wavelengths, while the bursts
with a positive drift cannot be identified in the spectrum at frequencies
higher than 350 MHz. Negative and positive frequency drifts are usually 
interpreted as electron beams travelling up and down in the corona, 
respectively. Some correlation can be found between these bursts and 
the count-rate peaks recorded by RHESSI in hard X-rays 
(Figure \ref{fig:11}). Since hard X-rays are usually created by downward 
directed electron beams when they hit denser regions, the type III
bursts with a positive frequency drift could be tracing these beams. 
The type III bursts could be imaged at 327, 236, 164, and 150 MHz 
by NRH, and the locations along one type III burst are shown in 
Figure \ref{fig:11}. The spatial separation between the RHESSI X-ray 
burst and the type III burst trajectory indicate that either the
sources were separate or the field lines were twisted. 
However, since the halo CME originated from this same active region,  
the lateral expansion of the CME could have affected the field lines. 
We also note the vicinity of the open field lines on the East side 
of the active region (shown in Figures \ref{fig:6} and \ref{fig:11}).

\begin{figure}
\centering
  \includegraphics[width=1.0\textwidth]{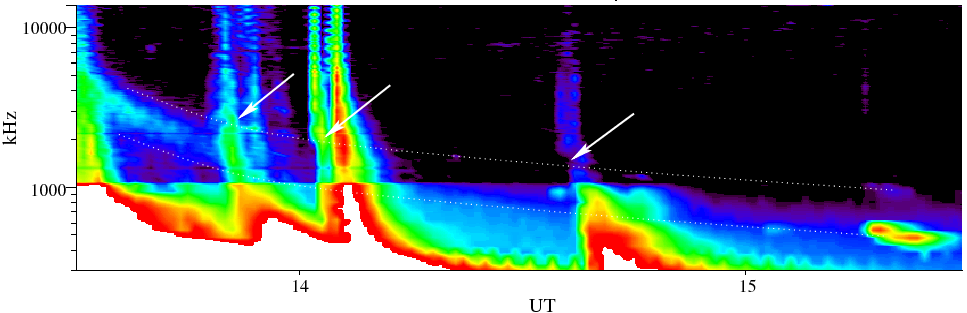}
\caption{{\it Wind}/WAVES RAD1+RAD2 spectrum at 300\,--\,14000 kHz at 
13:30\,--\,15:30 UT. The IP type II burst is visible as a patchy, broadband 
emission lane also after 14:00 UT (band edges outlined with dottted lines). 
The type III burst groups show a ``tilt'' at the time of the type II 
burst passage (times marked with arrows).}
\label{fig:12}       
\end{figure}

All the observed type III bursts continued to DH wavelengths, and the 
burst lanes look ``tilted'' at the time when they cross the type II burst 
lane, see Figure \ref{fig:12}. For the 13:48 UT type III burst group 
the tilt is not so obvious, but there is a brightening in the burst 
emission. For the 14:05 and 14:40 UT burst groups the type III tilt 
is more pronounced. It should be noted that in between 14:05 and 14:40 UT,
{\it i.e.}, in the time frame for the estimated proton acceleration, 
no other burst activity was observed in the dm--DH spectra. 

\section{Radio Source Heights and Velocities}

The radio emission during the flare\,--\,CME event showed several 
frequency-drifting structures. As plasma emission is directly related 
to electron density, $f_p$=9000$\sqrt{n_e}$ ($n_e$ in cm$^{-3}$ 
and $f_p$ in Hz), and electron density can be converted to atmospheric 
height with atmospheric density models, we calculate estimates 
for the radio source heights. These heights and deduced speeds need 
to be considered critically,  as they depend strongly on the selected
density models. Selection of model should be based on the knowledge 
of coronal conditions in the propagation region, but the available
information is often insufficient, see \inlinecite{pohjolainen07}.
Standard coronal density models such as \inlinecite{saito70} and 
\inlinecite{newkirk61}, and ``hybrid'' models such as 
\inlinecite{vrsnak04}, which is a mixture of the five-fold Saito model for 
coronal densities and the \inlinecite{leblanc98} model for IP densities, 
can be used to describe the undisturbed high corona at frequencies 
lower than $\approx$200 MHz (corresponding electron density 
$n_e$ $\lesssim$ 4$\times$10$^8$ cm$^{-3}$). 
For the high-frequency emission at 500\,--\,400 MHz (corresponding electron 
density $n_e$ $\approx$ 2--3$\times$10$^9$ cm$^{-3}$), we need 
high-density models such as the ten-fold Saito or a scale height method,
to describe densities in active region loops or streamer regions.
However, if the exiter of emission is propagating in the wake of earlier 
transients, in disturbed medium, the height estimates may not give
reliable results.  

The 300\,--\,100 MHz frequency range of the burst envelope of the 
narrow-band fluctuations (nbf) suggests a relatively large height 
for the emission source already at 13:09 UT. The estimated heliocentric 
distance at 100 MHz is 1.13\,--\,1.56 R$_\odot$, but the actual height value 
depends strongly on the density model used. From the radio spectrum in
Figure \ref{fig:10} it is not clear if the fluctuating emission near 
100 MHz is plasma emission at the fundamental frequency or at the second 
harmonic. Fundamental emission at 50 MHz would place the source height 
even higher in the corona. 
 
The drifting emission feature (df), observed at 400 MHz around 13:15 UT, 
had a height of 1.08 R$_\odot$ according to the 10-fold Saito density 
model. At 13:17 UT the emission had drifted to 300 MHz, which corresponds 
to a height of 1.14 R$_\odot$ using the same model. The deduced burst 
driver speed is approximately 500 km s$^{-1}$. The speed values 
calculated directly from the atmospheric models assume that the source 
motion is directed radially against the density gradient, and the 
speed has to be multiplied by 1/cos($\theta$) if these directions 
are separated by an angle $\theta$. Usually 45$^{\circ}$ and 
multiplication by 1.4 is considered as the realistic maximum 
correction needed. The scale height method, using the observed 
$\approx$0.8 MHz s$^{-1}$ frequency drift rate, yields a higher speed 
around 1420 km s$^{-1}$. We note that the burst sources of this feature, 
observed in the NRH images (shown in Figure \ref{fig:9}) suggest a 
projected source velocity of 1200\,--\,1400 km s$^{-1}$, which is close
to the value obtained with the scale height method.    
  
The DH type II burst, observed from 14 MHz at 13:23 UT to 2 MHz
at 13:47 UT, moved from a heliocentric height of 2.78 R$_\odot$ to
7.07 R$_\odot$ (Vr\v{s}nak, Magdaleni\'{c}, and Zlopec hybrid 
density model). The average speed along the burst lane was thus 
about 2070 km s$^{-1}$. Using the Saito
density model, instead of the hybrid, the heights are 2.07 
and 5.18 R$_\odot$, with a speed of 1500 km s$^{-1}$. 
In white light, the halo CME front was observed at a height of 
5.89 R$_\odot$ at 13:42 UT, which is closer to the heights
calculated with the hybrid density model.   

The observed start frequency of the second type III burst group, 
around 60 MHz near 13:48 UT, suggests that electrons were 
accelerated near heliocentric distances 1.3\,--\,1.8 R$_\odot$ (with all the
above-mentioned density models considered). The third type III group 
started at a much higher frequency, near 250 MHz at 14:05 UT, 
with bursts that showed both positive and negative frequency drifts. 
This high frequency indicates a low heliocentric height for 
acceleration, $<$1.2 R$_\odot$, with all the density models 
considered.

\section{Results}

The sequence of the observed features is listed in Table \ref{tbl1}.
The narrow-band fluctuations inside the drifting envelope started at 
13:09 UT, which is during the flare impulsive phase and also 
near the time of the estimated halo CME launch time. Radio 
pulsations can be due to magnetic reconnection, and recently
\inlinecite{kliem00} have associated this with plasmoid formation. 
The projected radio sources of this emission were above the active 
region, which is where the halo CME most probably originated. 
The source heights for the fluctuations are uncertain, since the 
observed plasma emission could also reflect the density of the dense 
active region loops,  along which beams of particles were 
travelling. As the bursts within the envelope had both positive and
negative frequency drifts, it is possible that magnetic trapping 
was present.

The drifting radio feature started next, at 13:14 UT near 500 MHz, 
which suggests a lower height in the atmosphere compared to 
the earlier narrow band fluctuations. The estimated speed for the 
burst driver is difficult to calculate because of the high starting 
frequency (high frequencies and high densities require that 
the standard atmospheric models have to be multiplied by arbitrary
adjustment factors). From the frequency drift and the density scale 
height we can make a rough estimate that the speed was around 
1420 km s$^{-1}$.
In the radio images the projected source speed was similar, 
1200\,--\,1400 km s$^{-1}$. The source trajectory suggests that this feature 
could have been associated with the EIT wave. The EIT wave features 
appeared near the locations of the drifting burst sources at 13:17 UT, 
near the bright EUV ``rim'' shown in Figures \ref{fig:7} and \ref{fig:9}. 
Unfortunately we could not calculate the speed of the EIT wave, as it was 
observed very near the solar limb. The halo CME speed was much higher 
than the estimated speed of this radio source, and the main halo-CME 
loop was also moving more to the West compared to the motion 
of the radio source. The radio source could also be showing a filament 
trajectory, as the observed EUV ribbons indicated a filament eruption.

\begin{table}
\caption[tbl1]{Summary of observations related to the 9 November 2002 event}
\label{tbl1}
\begin{tabular}{lll}
\hline
{\bf Time (UT)} &  {\bf Event} & {\bf Comments} \\
\hline
09:48\,--\,12:24& Activity near AR 0175 and AR 0177& loop openings and brightenings\\
                &                                  & near the northwestern limb\\ 
11:06           & Northwestern CME observed        & bright front in the Northwest,\\
                &                                  & CME speed 530 km s$^{-1}$\\
13:08           & M4.6 flare onset in AR 0180      & GOES, RHESSI\\ 
13:09           & Narrow-band fluctuations start   & frequency range 100\,--\,300 MHz\\
13:14           & Drifting radio feature starts    & start frequency $\approx$500 MHz\\
13:20           & DH type II radio burst observed  & at 14 MHz (height 2\,--\,3 R$_{\odot}$)\\
13:23           & Flare maximum                    & GOES \\
13:23\,--\,13:39& Electron release time            & if path length $>$2.1 AU \\
13:25           & EIT wave observed                & bright ``rim'' south of AR 0180\\
13:28\,--\,13:31& Type III burst group             & start frequency $<$60 MHz \\
13:31           & Halo CME observed at 4.2 R$_{\odot}$ & bright front in the Southwest\\
                &                                  & CME speed 1840 km s$^{-1}$\\
13:40\,--\,14:03& Electron release time            & if path length 1.0\,--\,1.4 AU \\
13:48\,--\,13:58& Type III burst group             & start frequency $\approx$60 MHz\\
13:57\,--\,14:27& Proton release time              & if path length 1.0\,--\,1.4 AU \\
14:02\,--\,14:06& Type III burst group             & start frequency $\approx$250 MHz\\
14:05           & RHESSI X-ray burst               & after satellite night-time\\
14:12           & Deepest EUV dimming              & \\
\hline
\end{tabular}
\end{table}

The DH type II burst that we were able to follow from 14 MHz down 
to 2 MHz, after which it was still observed as patchy emission down 
to kilometer waves, looks to have a clear association with the CME front.
The estimated burst heights fit well with a CME bow shock scenario. 
The deduced type II burst driver speed at 13:23\,--\,13:47 UT 
($\approx$2070 km s$^{-1}$) is not too far off from the plane-of-the-sky CME 
leading front speed measured at 13:31\,--\,13:42 UT (1895 km s$^{-1}$; 
acceleration to 2000 km s$^{-1}$ near 30 R$_\odot$) or the measured CME 
lateral expansion speed, $\approx$2300 km s$^{-1}$ before 14:00 UT. 
The height--time profiles of the type II burst and the CME suggest 
that both were formed during the flare impulsive phase. A timing match 
between the flare impulsive phase and the CME acceleration phase have
been noted by, {\it e.g.}, \inlinecite{zhang01}. 
 
The onset time analysis for protons gives the most probable acceleration 
time around 13:57\,--\,14:27 UT. The best candidate for the accelerator 
is the propagating CME bow shock, visible as the IP type II burst. 
Near 14:00 UT when the protons could have been released, the (projected) 
white-light CME front was located at height $\approx$9 R$_{\odot}$,
and the type II burst height was approximately 7\,--\,9 R$_{\odot}$
(actual number depending on the density model used). 
We emphasize that between 14:06 and 14:40 UT no other radio signatures 
were observed in the dynamic spectra.

The two type III burst groups that were observed to start at metric 
wavelengths (at 60 and 250 MHz) and continued to IP space 
suggest that electrons were accelerated at low heliocentric heights 
where they had access to open field lines. Low acceleration heights have 
been reported earlier by \inlinecite{klein05}, and their results of 
acceleration heights were between 0.1 and 0.5 R$_\odot$ above the photosphere 
agree with our estimates. They claim that acceleration usually happens 
well after the flare maximum.
The velocity dispersion analysis for the electron data, using the low 
path length obtained from the proton analysis, gives an estimate for 
the time of electron release at 13:40\,--\,14:03 UT. This time range agrees 
with our observations of the type III burst group observed to start 
at 13:48 UT. The start of the next type III burst group (14:02 UT) is 
somewhat late compared with the estimated electron injection time, 
but it is still possible. 
Acceleration happens well after the flare maximum (13:23 UT) 
in this case. But, if we accept the high path length obtained from 
the analysis of electron data only, the release time (13:23\,--\,13:39 UT) 
will match with the flare maximum.   

\inlinecite{krucker00} have suggested that accelerated electrons 
can be injected into the Earth-connected flux tube when a coronal 
transient (EIT wave) hits the field lines.
In our case we see the EIT wave passing the southwest ``rim'' 
approximately around 13:25\,--\,13:48 UT, but there is uncertainty 
due to the poor EIT image cadence. This time frame agrees with the 
estimated electron injection times with both the low and high 
path lengths and is close to the start time of the first type 
III burst group. However, since the origin of EUV emission is relatively 
low in the solar atmosphere compared to the observed radio 
emission of the first two type III burst groups, the association 
remains questionable. 
 
One LASCO image shows that the halo CME had a brightening on the side 
where the earlier, slower CME was propagating. This brightening 
appears between the available LASCO C2 images at 13:31 and 13:54 UT. 
Since the latitudinal and longitudinal distance between the two CME 
source regions was only 20\,--\,40$^{\circ}$, it is possible that the halo 
CME interacted with the earlier CME or with a streamer located in 
the northwest. A shock front could have formed in the interaction
region, due to compression. 
The radio enhancement was observed during 13:30\,--\,13:40 UT at 35\,--\,65 MHz,  
where the corresponding heliocentric height is around 1.5\,--\,2.0 R$_\odot$. 
At 13:54 UT the white-light brightening on the CME side had a height 
of $\approx$4 R$_\odot$. The deduced speed of the disturbance, if the radio 
feature and the white-light brigtening were connected, is approximately 
1100\,--\,1500 km s$^{-1}$.

\begin{figure}
\centering
  \includegraphics[width=1.0\textwidth]{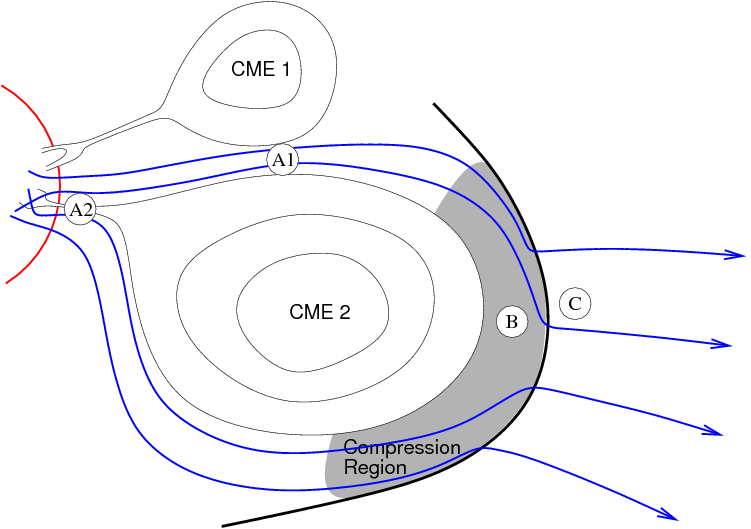}
\caption{A sketch of a possible geometry leading to the observed 
dynamics of the type III bursts: A1 and A2 are the possible 
reconnection sites where accelerated electrons may have 
access to open field lines; B is the region where the field 
lines traverse the compressed region behind the shock front 
driven by the CME; C is where the density and magnetic field 
decreases abruptly ahead of the shock, leading to a sudden drop 
of frequency and the focusing of the electron beam.}
\label{fig:13}       
\end{figure}

We also observed tilted type III burst lanes at times when the type 
IIIs passed the IP type II burst. After the tilt, some of the type 
III burst lanes brightened. We present in Figure \ref{fig:13} one possible 
geometry leading to the observed dynamics. Our sketch also explains the 
observed bidirectional motion of the metric type III bursts: \\
-- Electrons are accelerated in the possible reconnection site A2 in the
low corona, where they have access to open magnetic field lines 
(some open field lines were present in the potential magnetic field
maps prior to the event). 
The electron beams stream both upwards and downwards, as observed in 
the dynamic spectra and verified in NRH radio images.\\
-- Site A1 is also a possible reconnection site, because of 
the open field lines near the active region that produced the 
earlier CME1/earlier observed streamer region. Since A1 site is located 
higher in the solar atmosphere than site A2, A1 could be the acceleration 
region for the second type III burst group (start frequency near 60 MHz) 
and site A2 could be the the acceleration region for the later birectional 
type III burst group (start frequency near 250 MHz).\\      
-- Accelerated energetic electrons then enter the compression region B 
which is behind the shock front driven by CME2. The result is a bending 
of the type III burst lanes as the density gradient along the line 
of force is diminished. Particles propagate ahead of the shock wave, 
to region C, and experience a sudden decrease of the local density 
and magnetic field.  This in turn results in a drop of plasma frequency 
and focusing of the electron beam. The latter, in turn, leads to a more 
efficient generation of Langmuir waves and, hence, a re-brightening 
of the radio burst upstream of the shock front.

\section{Conclusions}

The release times for solar energetic particles, electrons and protons, 
can be obtained from, {\it e.g.}, velocity dispersion analysis. The 
feasibility of this method can be argued, as the path lengths can vary 
and there can be unknown propagation effects in the IP space. In our 
analysis protons in the 14\,--\,80 MeV energy range yield path lengths 
of 1.0\,--\,1.4 AU, while the analysed electrons suggest much higher 
path lengths of 2.4\,--\,2.7 AU. The path lengths of course have a 
strong effect on the estimated release times: with the low path 
lengths, electrons would be released just before the protons, and with 
adopting the high path lengths just for the electrons, electron 
release would have preceded proton release by more than half an hour.

Very high path lengths are usually taken as an indication of 
not-so-accurate timing results (Lintunen and Vainio, \citeyear{lintunen04};  
Saiz {\it et al.}, \citeyear{saiz05}). High path lengths can be the 
result of delayed onset observation on low-energy channels due to high 
pre-event backgrounds, high contrast between the spectral indices 
of pre-event background and the SEP event, prolonged injection 
of the particles, or interplanetary scattering of the particles. 
Therefore the release times yielded by the lower path lengths are
more plausible.  

We do not speculate on the idea of a ``seed population'' for the released
particles, although there is the possibility of injecting (flare) 
particles into the corona where the particles get accelerated later. 
The timing and the high start frequencies of the metric type III 
bursts that continue to the IP space suggest, however, that at least 
the energetic electrons are accelerated low in the corona and after 
the flare maximum.     

In summary, from the estimated release times for the electrons and 
protons compared with the multi-wavelength signatures of flare and coronal
mass ejection processes, we conclude that electrons were most probably 
accelerated at low coronal heights in the wake of a fast halo-type CME -- 
the source region could be reconnecting structures either on the side 
of the CME or near its legs -- and that the best candidate for proton 
accelerator is the CME bow shock, at a time when the CME was propagating 
high in the corona, above 9 R$_{\odot}$.  

The tilts and brightenings in the type III bursts at the times when 
they pass the IP type II burst is a new observation. We have presented 
a possible geometry leading to the observed dynamics, and point out 
that the tilted type III bursts can reveal propagating shock fronts 
in cases when the type II burst lanes are not visible in the spectrum.

\begin{acknowledgements}
We thank the radio group at LESIA, Observatoire de Paris, France, for the use 
of Nan\c{c}ay Radioheliograph and Decameter Array data, and 
the SOHO/EPHIN instrument team and the {\it Wind}/3DP instrument team  
and the Coordinated Data Analysis Web (CDAWeb) for providing the particle 
data. K-L. Klein and R. G\'{o}mez-Herrero are thanked for fruitful 
discussions. The comments and suggestions made by the anonymous referees
improved the paper significantly. The LASCO CME Catalog is generated 
and maintained by NASA and Catholic University of America in co-operation 
with the Naval Research laboratory. SOHO is an international co-operation 
project between ESA and NASA. 
N.J.L. wishes to thank the V\"ais\"al\"a Foundation of the 
Finnish Academy of Science and Letters for a travel grant to SPM-11. 
The Academy of Finland is thanked for partial support under project 
n:o 104329.
\end{acknowledgements}


\end{article}
\end{document}